\documentclass[aps,onecolumn,groupedaddress,nofootinbib,showpacs]{revtex4}

\usepackage{braket}
\usepackage{amsmath}
\usepackage{bm}
\usepackage{mathbbol}

\newcommand{\be}{\begin{equation}}
\newcommand{\ee}{\end{equation}}
\newcommand{\bea}{\begin{eqnarray}}
\newcommand{\eea}{\end{eqnarray}}

\newcommand{\p}{\partial}

\newcommand{\transp}{^{\text{T}}}
\newcommand{\inv}{^{-1}}
\newcommand{\dd}{\mathrm{d}}

\begin{document}

\title{Adiabatic motion and statistical mechanics via mass zero constrained dynamics}

\author{Sara Bonella,$^{\ast}$\textit{$^{a}$} Alessandro Coretti,\textit{$^{a,b}$} Rodolphe Vuilleumier,\textit{$^{c}$}and Giovanni Ciccotti\textit{$^{d,e,f}$}}

\affiliation{\textit{$^{a}$}~Centre Europ\'een de Calcul Atomique et Mol\'eculaire (CECAM), \'Ecole Polytechnique F\'ed\'erale de Lausanne, Batochime, Avenue Forel 2, 1015 Lausanne, Switzerland. Fax: 0041 21 693 19 85; Tel: 0041 21 693 19 79; E-mail: sara.bonella@epfl.ch \\
\textit{$^{b}$}~Department of Mathematical Sciences, Politecnico di Torino, Corso Duca degli Abruzzi 24, I-10129 Torino, Italy. \\
\textit{$^{c}$}~PASTEUR, D\'epartement de chimie, \'Ecole normale sup\'erieure, PSL University, Sorbonne Universit\'e, CNRS, 75005 Paris, France. \\
\textit{$^{d}$}~Institute for Applied Computing ``Mauro Picone'' (IAC), CNR Via dei Taurini 19, 00185 Rome, Italy. \\
\textit{$^{e}$}~Universit\`a di Roma La Sapienza, Ple. A. Moro 5, 00185 Roma, Italy. \\
\textit{$^{f}$}~School of Physics, University College of Dublin UCD-Belfield, Dublin 4, Ireland.
}
\begin{abstract}
In recent work [Coretti \textit{et al., The Journal of Chemical Physics}, 2018, \textbf{149}, 191102], a new algorithm to solve numerically the dynamics of the shell model for polarization was presented. The approach, broadly applicable to systems involving adiabatically separated dynamical variables, employs constrained molecular dynamics to strictly enforce the condition that the force on the fast degrees of freedom, modeled as having zero mass, is null at each time step. The algorithm is symplectic and fully time reversible, and results in stable and efficient propagation. In this paper we complete the discussion of the mechanics of mass zero constrained dynamics by showing how to adapt it to problems where the fast degrees of freedom must satisfy additional conditions. This extension includes, in particular, the important case of \textit{first principles} molecular dynamics. We then consider the statistical mechanics of the mass zero constrained dynamical system demonstrating that the marginal probability sampled by the dynamics in the physical phase space recovers the form of the Born-Oppenheimer probability density.
\end{abstract}

\maketitle

%
%
%




\section{Introduction}
In this paper, we reconsider the simulation of adiabatically separated systems using the recently introduced mass zero constrained molecular dynamics (MD) approach~\cite{coretti:2018b} by first extending the dynamics to the case of kinematically constrained mass zero evolution and then determining the statistical mechanics associated to the dynamical system. The inclusion of additional constraints enables to apply the method to new problems including \textit{first principles} molecular dynamics, while the statistical analysis defines the adiabatic probability for the physical system under consideration.

In adiabatically separated systems, the evolution of a set of physical degrees of freedom (dofs) is coupled to that of a separate, auxiliary, set which is usually introduced to account for some specific physics or chemistry. The mass of the auxiliary dofs is assumed much smaller than the mass(es) associated to the physical variables, and the adiabatic regime is fully achieved in the limit of null auxiliary mass. Due to the time-scale separation of the motion of the two sets of dofs, in this regime the physical variables evolve in a potential computed at the minimum with respect to the auxiliary variables. Depending on the specific system, additional (kinematic) conditions, such as orthonormality or sum rules, may have to be imposed on the auxiliary dofs, affecting the minimum search. In MD simulations, the choice of the algorithm to find the (free or conditioned) minimum is delicate. Typical schemes adopted so far include iterative methods, such as conjugate-gradient~\cite{aguado:2003a,jahn:2004}, extended Lagrangian schemes \`a la Car-Parrinello in which a small but finite mass is assigned to the auxiliary variables~\cite{sprik:1988,wilson:1993} or, more recently, alternative \textit{ad hoc} dynamics for the auxiliary variables like the so-called always stable predictor corrector approach~\cite{kolafa:2004,genzer:2004}. These methods, however, suffer from practical or conceptual limitations. Conjugate-gradient minimization is guaranteed to converge only in the case of a quadratic function to be minimized, and, for the general minimization problems typically associated with realistic condensed phase models, can be unstable~\cite{pacaud:2018} or expensive~\cite{pounds:2009} to fully converge. In Car-Parrinello propagation, it can be difficult or impossible to maintain adiabatic separation between the physical and auxiliary variables for sufficiently long simulation times~\cite{aguado:2003a} and the algorithm requires a very small timestep to integrate accurately the dynamics of the auxiliary variables. The always stable predictor corrector scheme is only approximately time reversible~\cite{kolafa:2004,genzer:2004} leading to energy drifts that are usually quenched via a Berendsen thermostat (thus raising questions on the ensemble sampled by the dynamics), and it contains system dependent parameters that can only be determined by trial and error. In spite of several interesting and successful applications of the approaches mentioned above, these limitations justify efforts to develop new algorithms for adiabatic dynamics.

In recent work~\cite{coretti:2018b}, we proposed to enforce the minimum energy condition using constrained molecular dynamics~\cite{ciccotti:1986}. In this approach, evolution equations for the physical and auxiliary degrees of freedom are derived under the constraint that the derivative of the Hamiltonian with respect to the auxiliary variables is zero along the trajectory of the physical degrees of freedom. Consistent with the adiabatic prescription, this is achieved by writing the method of constraints in the special case of mass zero constrained dofs~\cite{ryckaert:1981}. The resulting dynamical system is then integrated combining Verlet propagation with the SHAKE algorithm~\cite{ryckaert:1977} to satisfy the constraint, thus ensuring that the overall numerical integration of the system is symplectic and time reversible~\cite{leimkuhler:1994}. These formal properties guarantee exact energy conservation and stability of the evolution on the time-scales and with the time-step of standard MD for the physical degrees of freedom without the use of thermostats. Furthermore, the SHAKE algorithm, which requires iterations to solve the constraint equation, usually converges very fast in particular in non-linear problems, enabling to fulfill the minimum condition at the level of the numerical precision limit at an affordable cost. Importantly, SHAKE also prevents propagation of the error in each minimization along the trajectory. Proof of principle calculations~\cite{coretti:2018b} on a classical model for polarizable systems have indeed shown that the mass zero constrained dynamics is an interesting and cost-effective alternative to currently adopted schemes.

In this paper, we extend the approach to models where the fast degrees of freedom must satisfy additional conditions, such as specific normalization or orthogonality requirements. This development enables, in particular, to adapt the method to the very interesting case of \textit{first principles} molecular dynamics, but is relevant also for classical polarizable models, for example in connection with the global electroneutrality constraint in simulation of capacitors. The similarities between classical polarizable models and \textit{first principles} molecular dynamics have been long recognized, and algorithms have often migrated from one area to the other, a notable example being the seminal contributions of Sprik et al.~\cite{sprik:1988}. Here we show that these similarities fully persist within the mass zero constrained approach. Having generalized the mechanics of the system, we consider the question of the statistical mechanics associated with mass zero constrained dynamics. As mentioned above, this approach entails an extended phase space in which both the physical and the auxiliary variables are treated as dynamical degrees of freedom. In the following, we derive the probability density sampled in this extended phase space, focusing in particular, on the marginal in the phase space of the physical variables, thus identifying unambiguously the full adiabatic limit for the probability associated to these degrees of freedom.


The paper is organized as follows. For the reader's convenience, in section~\ref{sec:Mechanics} we begin by summarizing the derivation of the mass zero constrained dynamics for the case of classical polarization of shell-like models where only the minimization of the potential is required. We then show, using orbital-free Density Functional Theory~\cite{pearson:1993, ligneres:2005-inbook} as a simple but significative illustrative example, how to extend the dynamical system when an additional condition, in this case the conservation of the number of electrons, must be enforced. The derivation we present can be generalized easily to situations, such as Kohn-Sham orbitals~\cite{kohn:1965,marx:2012-book},  in which more than one additional condition must be satisfied. In section~\ref{sec:Statistics}, we discuss the statistical mechanics of mass zero constrained dynamics showing that the marginal probability density in the physical degrees of freedom recovers the form usually assumed for the Born-Oppenheimer probability.


\section{Adiabatic dynamics via the mass zero constrained evolution}\label{sec:Mechanics}

\subsection{Unconditioned minimization}\label{sec:ClassicalPolarizableModelsUnconditioned}
To provide a specific example for the mass zero constrained approach in the case of simple minimization of the potential with respect to the fast variables, we shall illustrate our (general) formalism via the case of classical polarizable models. These models are crucial in the simulation of ionic systems of theoretical and technological interest such as, for example, energy storage devices (batteries and supercapacitors)~\cite{simon:2008,armand:2008,beguin:2014}. 
An accurate description of polarization is essential~\cite{stoneham:1986}, for example, to compute with sufficient precision phonon dispersion curves in crystals or transport and structural properties in fluids, and to account for some apparently anomalous ordered structures observed in solids and melts without the need to invoke charge transfer effects. \textit{First principles} MD can, in principle, include these effects directly via the quantum treatment of the electronic charge distribution. However, since the sizes and time-scales necessary to compute the properties of ionic systems are substantial, semi-empirical potentials are still the most convenient modeling tool in this field. In these potentials, appropriate sets of auxiliary dynamical variables mimic changes in the electronic charge density. These variables might appear in the terms of the potential representing short-range repulsion, Coulomb and dispersion interactions, and their coupling to the ionic dofs is controlled by parameters obtained from comparisons with experiments or with accurate ab initio calculations. One of the first examples of this kind was the shell model~\cite{dick:1958,jacucci:1975,jacucci:1976}, which accounts for dipole polarization, followed by improvements taking into account interactions due to quadrupoles~\cite{wilson:1996a} and changes in the ions size and shape~\cite{wilson:1996b,rowley:1998}. More recently, models for capacitors have been proposed that include the mutual polarization of the elements combining a multipole description of the electrolyte with the so called fluctuating charge model~\cite{sprik:1988} for the electrodes.

Let us indicate with $R_i$ ($i=1,...,3N$) the coordinates associated with the $N$ physical variables in the system, and with $s_{\alpha}$ ($\alpha=1,...,M$) the $M$ auxiliary degrees of freedom introduced to represent specific polarization effects. The $s_{\alpha}$ variables may represent positions (as in the shell model) or different types of degrees of freedom (e.g. dipoles or quadrupoles, or charges) and their physical dimensions and number vary accordingly. The interaction potential of the system has the generic form $V(R, s)$, where we have used the notation $R=\{R_1,...,R_{3N}\}$ and $s=\{s_1,...,s_M\}$. The dynamics of the system is defined as follows. The auxiliary variables, having zero mass, adapt instantaneously to the position of the physical degrees of freedom so as to satisfy

\begin{equation}\label{eq:ConstraintConditions}
\sigma_{\alpha}(R, s) \equiv  \frac{\partial V(R, s)}{\partial s_\alpha} = 0 \qquad \alpha = 1,\dots,M
\end{equation}
In the following, we indicate with $\hat{s}$ the values of the auxiliary variables satisfying the conditions above. Note that, due to the dependence of the potential on the physical degrees of freedom, $\hat{s}=\hat{s}(R)$. The evolution of the physical variables, of mass $m_i$, is then given by

\begin{equation}\label{eq:BODynamicsPolar}
m_i\ddot{R}_i = -\left [\frac{\p V(R, s)}{\p R_i}\right]_{s=\hat{s}} 
\end{equation}
The key idea of the mass zero constrained dynamics is to interpret eq.~\ref{eq:ConstraintConditions} as a set of $M$ holonomic constraints and derive the evolution equations for the overall constrained system. This is done most conveniently by assigning at first a finite mass $\mu$ to the auxiliary variables and considering the extended Lagrangian:
\begin{equation}\label{eq:MaZeLagrangian}
L(R, \dot{R}, s, \dot{s}) = \frac{1}{2}\sum_{i=1}^{3N}m_i\dot{R}^{2}_{i} + \frac{1}{2}\sum_{\alpha=1}^M\mu\dot{s}^{2}_{\alpha} - V(R, s)
\end{equation}
together with eq.~\ref{eq:ConstraintConditions} to obtain the constrained evolution equations

\begin{equation}
\begin{aligned}
m_i\ddot{R}_i &= -\frac{\p V(R, s)}{\p R_i}  - \sum_{\beta = 1}^M \lambda_{\beta} \frac{\p \sigma_{\beta}(R, s)}{\p R_i}\\
\mu\ddot{s}_\alpha &= -\frac{\p V(R, s)}{\p s_{\alpha}} - \sum_{\beta = 1}^M \lambda_{\beta} \frac{\p \sigma_{\beta}(R, s)}{\p s_{\alpha}}
\end{aligned}
\end{equation}
where the $\lambda_{\beta}$ are the Lagrange multipliers associated with the constraints. The system above can be simplified by observing that the gradient of the potential in the second equation is null due to the minimum condition. The equation for the auxiliary degrees of freedom can then be rearranged as
\begin{equation}
\ddot{s}_\alpha = - \sum_{\beta = 1}^M \frac{\lambda_{\beta}}{\mu} \frac{\p \sigma_{\beta}(R, s)}{\p s_{\alpha}}
\end{equation}
We now consider the limit $\mu\rightarrow 0$. The equation above shows that, in order for the auxiliary variables to have finite acceleration, the ratio $\gamma_\alpha = \lim_{\mu\to0}\frac{\lambda_\alpha}{\mu}$ must remain finite. This implies that the Lagrange multipliers $\lambda_\alpha$ are proportional to $\mu$. In the limit then, the constraint forces on the physical degrees of freedom vanish and we have
\begin{equation}\label{eq:MaZeEquations}
\begin{aligned}
m_i\ddot{R}_i &= -\frac{\p V(R, s)}{\p R_i} \\ 
\ddot{s}_\alpha &= -\sum_{\beta=1}^M\gamma_\beta\frac{\p \sigma_\beta(R, s)}{\p s_\alpha}
\end{aligned}
\end{equation}
Eq.~\ref{eq:MaZeEquations} is the main result of ref.~\cite{coretti:2018b} and defines the mass zero constrained dynamics. It shows that, consistent with eq.~\ref{eq:BODynamicsPolar}, the evolution of the physical variables does not depend directly on the constraints. Due to the constraints, on the other hand, the (free) $s$ variables satisfy the minimum condition. This implies that this condition is automatically satisfied also in the first equation which is then equivalent to eq.~\ref{eq:BODynamicsPolar}. The system above is an exact, classical, evolution for all degrees of freedom that rigorously enforces the zero mass limit, and therefore leads to full adiabatic separation, for the auxiliary variables. The numerical integration of the first equation can be performed with any standard MD algorithms (e.g. Verlet) with a time-step determined only by the explicit physical force term. In addition, at each time-step, the Lagrange multipliers  $\gamma_\alpha$, that appear as unknown, time-dependent parameters in the dynamical system, must be determined. This is done enforcing the constraint, $\sigma_{\alpha}(R(t+dt), s(t+dt))=0$, at the position predicted by the MD algorithm as described in~\cite{ryckaert:1977,ciccotti:1986}. In this way propagation of the error is prevented. In current implementations of the approach, the constraint is satisfied via the SHAKE iterative algorithm, which was proven to be symplectic and time reversible~\cite{leimkuhler:1994,leimkuhler:2004-book}. 

\subsection{Conditioned minimization}
So far, we have assumed that the only conditions to be satisfied by the auxiliary dofs were those expressed by eq.~\ref{eq:ConstraintConditions}. While this is often the case, e.g. the shell model or classical treatment of multipoles, there are important problems in which additional requirements are placed on these degrees of freedom. The most interesting example is probably given by \textit{first principles} MD, that we shall therefore adopt hereafter to illustrate how the mass zero constrained dynamics can be extended to include generic additional conditions. In \textit{first principles} MD, classical propagation of the nuclei is combined with a quantum evaluation of the forces, obtained on-the-fly from accurate electronic structure calculations. In this approach, adiabatic separation of the nuclear and electronic dofs is invoked to claim that, at each nuclear configuration, the electronic subsystem has relaxed in its ground state. The ground state energy is found by minimizing the quantum expectation value of the electronic Hamiltonian with respect to the (normalized) electronic state. This energy is a functional of the electronic dofs that depends parametrically on the nuclear positions, thus providing the force on the nuclei via the appropriate derivatives.
Currently, the best compromise between cost and accuracy in determining the ground state energy is offered by Density Functional Theory (DFT). DFT-\textit{First principles} MD has indeed seen an impressive growth since the mid 1980s and is adopted in areas ranging from materials modeling to drug design~\cite{cavalli:2006,briddon:2011,marx:2012-book}. The success of available methods, however, still leaves room for methodological and algorithmic improvements to secure stable, unbiased, and efficient long-time evolution as witnessed by several recent efforts~\cite{niklasson:2006,niklasson:2007,kuhne:2007,niklasson:2008,niklasson:2009,niklasson:2012,lin:2014,kuhne:2014,niklasson:2017,albaugh:2018}.
To demonstrate how mass zero constrained MD can be extended to this problem, we specify our discussion to the so-called orbital-free DFT scheme~\cite{pearson:1993, ligneres:2005-inbook} in which the electronic state is represented directly via the electronic density, $n(r)$. We have chosen this representation because it permits to illustrate the key ingredients of our approach with minimal notational complexity since it requires to include only one additional condition, i.e. conservation of the number or electrons, to the minimization. The extension to the case of multiple conditions is straightforward and will be discussed at the end of the section. 
The electronic density is a field in Cartesian space, but in the following we shall restrict our considerations to an appropriate discretization of this quantity. This discretization must, of course, always be enforced in numerical implementations and can be realized via a decomposition on a basis set or a representation on a grid. Below, we shall consider a grid decomposition but the discussion can be similarly performed in a basis. Let us then indicate with  $r_k$ the points on the grid and use the notation $n(r_k)\equiv n_k$ ($k=1,\dots,M$) for the discretized density. These $n_k$ variables play the role of the auxiliary dofs $s_\alpha$ of the previous section. The Born-Oppenheimer evolution equations for the nuclear degrees of freedom are given by
\begin{equation}
m_i\ddot{R}_i = -\left [\frac{\p E_0(R, n)}{\p R_i}\right]_{n=\hat{n}}
\end{equation}
where $E_0(R, n)$ is the ground state expectation value of the electronic Hamiltonian, $\bra{\psi_0}\mathcal{H}_{\text{el}}\ket{\psi_0}$, expressed as a function of the discretized density variables $n=\{n_1,\dots,n_M\}$. In the equation above, we have indicated with $\hat{n}$ the values of these variables that minimize $E_0(R,n)$ subject to the condition
\begin{equation}
\label{eq:Nel}
 f(n)= \sum_{k=1}^Mn_k \delta V - N_{\text{el}} = 0
\end{equation}
which expresses, in discrete form, the fact that the integral of the electronic density must be equal to the total number of electrons, $ N_{\text{el}}$, in the system ($\delta V$ is the elementary volume). Eq.~\ref{eq:Nel} implies that not all variations of the $n_k$ are independent, and this must be accounted for in the statement of the minimum condition. This can be done most conveniently via the Lagrange multiplier method by introducing the auxiliary function
\begin{equation}
\mathcal{E}(R, n, \nu)\equiv E_0(R, n)-\nu f(n)
\end{equation}
where $\nu$ is the Lagrange multiplier associated to the condition above. The solution $\hat{n}$ for $n$ satisfying eq.~\ref{eq:Nel} is then given by the stationary point $(\hat{n},\hat{\nu})$ of $\mathcal{E}(R, n, \nu)$.\cite{lanczos:1970-book, allaire_numerical_2007} This leads to the $M+1$ conditions
\begin{equation}
\label{eq:SigmaNkPiusNu}
\begin{aligned}
\sigma_k(R, n, \nu) &= \frac{\partial \mathcal{E}(R, n, \nu)}{\partial n_k} = 0  \,\,\,\, (k=1,\dots,M) \\
\sigma_{M+1}(R, n, \nu) &= \frac{\partial \mathcal{E}(R, n, \nu)}{\partial \nu} = 0.
\end{aligned}
\end{equation}
Note that, as usual in the Lagrange multiplier scheme, the last equation above is in fact the additional condition eq.~\ref{eq:Nel} but now obtained as a result of an optimization problem in the space that includes the Lagrange multiplier $\nu$ as a variable.
Within this framework, we can proceed in analogy with the previous section by defining the following extended Lagrangian, which includes also $\nu$ as an auxiliary variable,
\begin{equation}
\label{eq:LNkPiusNu}
L(R, \dot{R}, n, \dot{n}, \nu, \dot{\nu}) = \frac{1}{2}\sum_{i=1}^{3N}m_i\dot{R}^{2}_{i} + \frac{1}{2}\mu_n\sum_{k=1}^M\dot{n}^2_k + \frac{1}{2}\mu_{\nu} \dot{\nu}^2- \mathcal{E}(R, n, \nu),
\end{equation}
and interpreting eq.~\ref{eq:SigmaNkPiusNu} as $M+1$ holonomic constraints on the system. In the equation above, we have introduced, for dimensional consistency, two finite masses $\mu_n$ and $\mu_{\nu}$ related to the auxiliary variables associated with the discretized density and with the Lagrange multiplier $\nu$, respectively. Proceeding as in the previous section, and imposing that the masses of both sets of auxiliary variables tend to zero, the mass zero constrained dynamical system is given by
\begin{equation}\label{eq:BODynamicsPlusNu}
\begin{aligned}
m_i\ddot{R}_i &= -\frac{\p E_0(R, n)}{\p R_i}, \\
\ddot{n}_i &= -\sum_{\alpha =1}^{M+1}\gamma_{\alpha} \frac{\partial \sigma_{\alpha}(R, n, \nu)}{\partial n_i}, \\
\ddot{\nu} &= -\sum_{\alpha =1}^{M+1}\gamma_{\alpha} \frac{\partial \sigma_{\alpha}(R, n, \nu)}{\partial \nu}
\end{aligned}
\end{equation}
where, in the first equation, we have used the fact that $\frac{\p \mathcal{E}(R, n, \nu)}{\p R_i}=\frac{\p E_0(R, n)}{\p R_i}$, while the first derivatives of $\mathcal{E}$ in the second and third line vanish due to the stationary point conditions. 
In this extended system, we now recover a situation similar to that of the polarizable models and the numerical integration of the equations above can again be performed by combining standard algorithms such as Verlet with SHAKE to enforce the constraints. 
In this respect, note that the potential energy term $\mathcal{E}(R, n, \nu)$ is linear in $\nu$ with the consequence that the second derivative of $\mathcal{E}$ is zero in this direction. Thus the matrix of second derivatives at the stationary point $(\hat{n},\hat{\nu})$ is non-positive definite and $(\hat{n},\hat{\nu})$ is a saddle point.\cite{lanczos:1970-book, allaire_numerical_2007} However, this does not hinder the procedure since SHAKE only requires non-singularity but not positive definiteness of the matrix of second-derivatives.

The framework described above is quite general and can be easily extended to the case of $G$ additional conditions, $f_{j}(s)$ (with $j=1,\dots,G$), for the auxiliary degrees of freedom. This can be done by introducing, and subsequently treating as an additional auxiliary variable, one Lagrange multiplier, $\nu_j$, per additional condition and modifying the Lagrangian by subtracting the sum $\sum_{j=1}^G\nu_j f_{j}(s)$ to the energy functional. This procedure enables, for example, to apply the mass zero dynamics procedure to Kohn-Sham orbital based \textit{first principle} MD thus further illustrating the overall interest of this approach.

\section{Statistical mechanics of the mass zero constrained evolution}\label{sec:Statistics}
We now investigate the probability density sampled by the dynamical system eq.~\ref{eq:MaZeEquations} . This is an interesting question because the use of constraints may induce a non-trivial metrics in the phase space of the system~\cite{tuckerman:2011} and appropriate reweighting of averaging might be necessary when computing statistical properties~\cite{ciccotti:2005} in the physical phase space. To set the stage for the discussion, and proceeding in analogy with~\cite{ciccotti:2018}, it is useful to introduce the change of variables (we avoid considering also the $\nu$ variable(s) for notational simplicity)
\begin{equation}\label{eq:ChangeOfVariables}
\begin{aligned}
R_i &\mapsto \rho_i = R_i \\
s_\alpha &\mapsto \sigma_\alpha = \sigma_\alpha(R, s)
\end{aligned}
\end{equation}
Observables computed in the new variables will be denoted in calligraphic font.
In the following, we shall also use the notation $\upsilon = (\rho, \sigma)$ where $\rho=\{\rho_1,...,\rho_{3N}\}$ and $\sigma=\{\sigma_1,...,\sigma_M\}$. The Lagrangian in the new set of coordinates is obtained from eq.~\ref{eq:MaZeLagrangian} as $\mathcal{L}(\upsilon,\dot{\upsilon})=L\bigl(R(\upsilon),s(\upsilon),\dot{R}(\upsilon,\dot{\upsilon}),\dot{s}(\upsilon,\dot{\upsilon})\bigr)$. The Hamiltonian of the system in the new coordinates is then derived via a standard Legendre transform and can be written as
\begin{equation}\label{eq:Hamiltonian}
\mathcal{H}(\upsilon, \pi^{\upsilon}) = \mathcal{K}(\upsilon,\pi^{\upsilon}) + \mathcal{V}(\upsilon) \equiv  \frac{1}{2}\pi^{\upsilon \transp} \mathbb{M}^{-1}(\upsilon)\pi^{\upsilon} + \mathcal{V}(\upsilon)
\end{equation}
where $\mathcal{K}(\upsilon,\pi^{\upsilon})$ is the kinetic energy in the new variables, and the momentum $\pi^{\upsilon}$ is given by
\begin{equation}
\pi^{\upsilon}_k = \frac{\p \mathcal{L}(\upsilon, \dot{\upsilon})}{\p \dot{\upsilon}_k} 
=
\begin{cases} 
\pi^{\rho}_i = \frac{\p \mathcal{L}(\rho, \dot{\rho}, \sigma, \dot{\sigma})}{\p \dot{\rho}_i} & \text{if $k = 1,\dots, 3N$}\\
\pi^{\sigma}_\alpha = \frac{\p \mathcal{L}(\rho, \dot{\rho}, \sigma, \dot{\sigma})}{\p \dot{\sigma}_\alpha} & \text{if $k = 3N+1,\dots, 3N+M$ }\\
\end{cases}
\end{equation}
(as always, $i=1,\dots,3N$ and $\alpha = 1,\dots,M$) and $\mathbb{M}^{-1}(\upsilon)$ is the inverse of the metric matrix associated with the new variables
\begin{equation}
\begin{aligned}
\mathbb{M}_{kk'} = \sum_{i=1}^{3N}m_i\frac{\p R_i}{\p \upsilon_k}\frac{\p R_i}{\p \upsilon_{k'}} + \sum_{\alpha=1}^{M}\mu\frac{\p s_\alpha}{\p \upsilon_k}\frac{\p s_\alpha}{\p \upsilon_{k'}}
\end{aligned}
\end{equation}
The metric matrix and its inverse can be conveniently expressed in block form as
\begin{equation}\label{eq:MassMatrix}
\mathbb{M} = 
\begin{bmatrix}
\mathbb{A} & \mathbb{B} \\
\mathbb{B}\transp & \mathbb{\Gamma}
\end{bmatrix}
\quad
\text{with inverse}
\quad
{\mathbb{M}}\inv = 
\begin{bmatrix}
\mathbb{\Delta} & \mathbb{E} \\
{\mathbb{E}}\transp & \mathbb{Z}
\end{bmatrix}
\end{equation}
where $\mathbb{A}$ and $\mathbb{\Delta}$ are $3N\times 3N$ matrices, $\mathbb{B}$ and $\mathbb{E}$ are $3N\times M$ matrices, and $\mathbb{\Gamma}$ and $\mathbb{Z}$ are $M\times M$ matrices, whose expressions are given in the Appendix. The Hamiltonian in eq.~\ref{eq:Hamiltonian} can be evaluated explicitly on the constrained hypersurface in the full phase space. To that end, it is important to recognize that, since the set of constraints in eq.~\ref{eq:ConstraintConditions} have to be fulfilled at all times, the additional conditions $\dot{\sigma}=0$ hold. The definition $\dot{\upsilon}={\mathbb{M}}\inv\pi^{\upsilon}$ then implies that, when the constraints are imposed, the momenta $\pi^{\sigma}$ must satisfy 
\begin{equation}\label{eq:ConstrainedMomentum}
\hat{\pi}^{\sigma} = -{\widetilde{\mathbb{Z}}}\inv{\widetilde{\mathbb{E}}}\transp\pi^{\rho}
\end{equation}
where the tildes indicate that all matrices are evaluated at $\sigma=0$. The constrained motion then occurs on the hypersurface $\sigma=0$, $\pi^{\sigma}=\hat{\pi}^{\sigma}$ and we have
\begin{equation}
\begin{aligned}
& \mathcal{H}(\rho, \sigma = 0, \pi^{\rho}, \pi^{\sigma} = \hat{\pi}^\sigma) \\
 & = \frac{1}{2}(\pi^{\rho}, \hat{\pi}^\sigma)
\begin{bmatrix}
\widetilde{\mathbb{\Delta}} & \widetilde{\mathbb{E}} \\
{\widetilde{\mathbb{E}}}\transp & \widetilde{\mathbb{Z}}
\end{bmatrix}
\begin{pmatrix}
\pi^{\rho} \\
\hat{\pi}^\sigma
\end{pmatrix}
+ \mathcal{V}(\rho, \sigma=0) \\
\end{aligned}
\end{equation}
The Hamiltonian can be written in a more suitable form by observing that the block expression of the product $\mathbb{M}\mathbb{M}\inv=\mathbb{1}$ imposes  $\widetilde{\mathbb{A}}\widetilde{\mathbb{\Delta}}+\widetilde{\mathbb{B}}\widetilde{\mathbb{E}}\transp=\mathbb{1}$ and $\widetilde{\mathbb{A}}\widetilde{\mathbb{E}}+\widetilde{\mathbb{B}}\widetilde{\mathbb{Z}}=\mathbb{0}$. These two relationships, in turn, imply $\widetilde{\mathbb{\Delta}} - \widetilde{\mathbb{E}} {\widetilde{\mathbb{Z}}}\inv{\widetilde{\mathbb{E}}}\transp=\widetilde{\mathbb{A}}\inv$ so that
\begin{equation}\label{eq:ConstrainedHamiltonian}
\begin{aligned}
& \mathcal{H}(\rho, \sigma = 0, \pi^{\rho}, \pi^{\sigma} = \hat{\pi}^\sigma)  \\
& =  \frac{1}{2}
[\pi^{\rho\transp}(\widetilde{\mathbb{\Delta}} - \widetilde{\mathbb{E}}{\widetilde{\mathbb{Z}}}\inv{\widetilde{\mathbb{E}}}\transp)\pi^{\rho}] + \mathcal{V}(\rho, \sigma=0) \\
& = \frac{1}{2}
\pi^{\rho\transp}\widetilde{\mathbb{A}}\inv\pi^{\rho} + \mathcal{V}(\rho, \sigma=0) 
\end{aligned} 
\end{equation}
The microcanonical partition function in the full phase space $(\rho,\sigma,\pi^{\rho},\pi^{\sigma})$ can now be written as 
\begin{equation}
\begin{aligned}
Z = \int&\dd^N\rho\dd^N\pi^{\rho}\dd^M\sigma\dd^M\pi^{\sigma} \delta^M(\sigma)\delta^M(\pi^{\sigma} - \hat{\pi}^{\sigma})\times \\
&\times \delta(\mathcal{H}(\rho, \sigma, \pi^{\rho}, \pi^{\sigma})-E)
\end{aligned}
\end{equation}
while the expression of the microcanonical average of observable $\mathcal{O}$ is given by
\begin{equation}
\begin{aligned}
\Braket{\mathcal{O}} = \frac{1}{Z}\int&\dd^N\rho\dd^N\pi^{\rho}\dd^M\sigma\dd^M\pi^{\sigma} \delta^M(\sigma)\delta^M(\pi^{\sigma} - \hat{\pi}^{\sigma})\\
&\times \mathcal{O}(\rho, \sigma, \pi^{\rho}, \pi^{\sigma})\delta(\mathcal{H}(\rho, \sigma, \pi^{\rho}, \pi^{\sigma})-E)\\
\end{aligned}
\end{equation}
The expression above can be usefully simplified by first integrating over the $\pi^{\sigma}$ variables
\begin{equation}
\begin{aligned}
\Braket{\mathcal{O}} = \frac{1}{Z'}\int&\dd^N\rho\dd^N\pi^{\rho}\dd^M\sigma \delta^M(\sigma)\\
&\times \mathcal{O}(\rho, \sigma, \pi^{\rho}, \hat{\pi}^{\sigma})\delta(\mathcal{H}(\rho, \sigma, \pi^{\rho}, \hat{\pi}^{\sigma})-E)\\
\end{aligned}
\end{equation}
and then performing the change of variables $\sigma_\alpha \mapsto s_\alpha$, $\rho_i \mapsto R_i$  to obtain at first
\begin{equation}
\begin{aligned}
\Braket{\mathcal{O}} = \frac{1}{Z'}\int&\dd^NR\dd^N\pi^{\rho}\dd^Ms |J(R)| \delta^M(\sigma(R,s))\\
&\times \mathcal{O}(R, \sigma(R,s), \pi^{\rho}, \hat{\pi}^{\sigma})\delta(\mathcal{H}(R, \sigma(R,s), \pi^{\rho}, \hat{\pi}^{\sigma})-E)\\
\end{aligned}
\end{equation}
where $|J(R)|$ is the Jacobian of the coordinate transformation, which reduces to $\det\left[\frac{\partial \sigma}{\partial s}\right]$. Then, making the dependence on $s$ of the delta explicit, we get
\begin{equation}
\begin{aligned}
\Braket{\mathcal{O}} =\frac{1}{Z'}\int&\dd^NR\dd^N\pi^{\rho}\dd^Ms |J(R)| |J(R)|^{-1} \delta^M(s-\hat{s}(R))\\
&\times O(R, s, \pi^{\rho}, \hat{\pi}^{\sigma})\delta(H(R, s, \pi^{\rho}, \hat{\pi}^{\sigma})-E)\\
\end{aligned}
\end{equation}
where we have $O(R, s, \pi^\rho, \hat{\pi}^\sigma) = \mathcal{O}(R, \sigma(R,s), \pi^{\rho}, \hat{\pi}^{\sigma})$.
In this last equality we have used the properties of the delta of a vector function of the integration variable to express the constraint condition directly as a function of the $s$, with $\hat{s}(R)$ such that $\sigma(R,\hat{s})=0$ (we assume, as commonly done in the Born-Oppenheimer framework that this expression has, for any $R$, a single root). Finally, performing the integral over the $s$ variables, and noting that the product of Jacobians in the integrand simplifies, we obtain
\begin{equation}
\begin{aligned}\label{eq:ZeroMassMargProb}
\Braket{\mathcal{O}} =\frac{1}{Z'}\int&\dd^N R\dd^N\pi^{\rho}O(R, \hat{s}, \pi^{\rho}, \hat{\pi}^{\sigma})\delta(H(R, \hat{s}, \pi^{\rho}, \hat{\pi}^{\sigma})-E)\\
\end{aligned}
\end{equation}
where (see eq.~\ref{eq:ConstrainedHamiltonian})
\begin{equation}
H(R, \hat{s}, \pi^{\rho}, \hat{\pi}^{\sigma})=\frac{1}{2}\pi^{\rho\transp}\widetilde{\mathbb{A}}\inv\pi^{\rho} + V(R, \hat{s})
\end{equation}
Let us now consider the limit $\mu\rightarrow 0$ of the expressions above. 
To set the stage, observe that, based on the definition in the Appendix, we can conveniently rewrite
\begin{equation}
\widetilde{\mathbb{A}}=\mathbb{D}+\mu\widetilde{\mathbb{R}}=\mathbb{D}\left[\mathbb{1}+\mu\mathbb{D}\inv\widetilde{\mathbb{R}}\right]
\end{equation}
where $\mathbb{D}_{jj'}=m_j\delta_{jj'}$ and $\widetilde{\mathbb{R}}_{jj'}=\sum_{\alpha=1}^M \frac{\p s_{\alpha}}{\p \rho_j} \frac{\p s_{\alpha}}{\p \rho_{j'}}$ so that
\begin{equation}\label{eq:AMatrix}
\widetilde{\mathbb{A}}\inv=\left[\mathbb{1}+\mu\mathbb{D}\inv\widetilde{\mathbb{R}}\right]\inv\mathbb{D}\inv
\end{equation} 
Furthermore, the relation $\pi^{\upsilon}={\mathbb{M}}\dot{\upsilon}$ implies, $\pi^{\rho}=\mathbb{A}\dot{\rho}+\mathbb{B}\dot{\sigma}$ giving, on the constrained hypersurface
\begin{equation}\label{eq:MomentaRel}
\pi^{\rho}=\widetilde{\mathbb{A}}\dot{\rho}=\widetilde{\mathbb{A}}\dot{R}
\end{equation}
where in the last equality we used $\dot{\rho}=\dot{R}$, as implied by eq.~\ref{eq:ChangeOfVariables}. Finally, from $\mathbb{M}\mathbb{M}\inv=\mathbb{1}$, the identities  $\widetilde{\mathbb{A}}\widetilde{\mathbb{E}}+\widetilde{\mathbb{B}}\widetilde{\mathbb{Z}}=\mathbb{0}$ and $\widetilde{\mathbb{B}}\transp\widetilde{\mathbb{E}}+\widetilde{\mathbb{\Gamma}}\widetilde{\mathbb{Z}}=\mathbb{1}$ follow for the involved submatrices. Using these identities, we obtain
\begin{equation}
\begin{aligned}
\widetilde{\mathbb{Z}}\inv&=\widetilde{\mathbb{\Gamma}}-\widetilde{\mathbb{B}}\transp\widetilde{\mathbb{A}}\inv\widetilde{\mathbb{B}} \\
\end{aligned}
\end{equation}
so that
\begin{equation}
\begin{aligned}
\hat{\pi}^{\sigma} &= -{\widetilde{\mathbb{Z}}}\inv{\widetilde{\mathbb{E}}}\transp\pi^{\rho} \\
&= \left[\widetilde{\mathbb{\Gamma}}-\widetilde{\mathbb{B}}\transp\widetilde{\mathbb{A}}\inv\widetilde{\mathbb{B}}\right]{\widetilde{\mathbb{E}}}\transp\pi^{\rho} 
\end{aligned}
\end{equation}
We can now take the limit $\mu \rightarrow 0$ by observing first that $\widetilde{\mathbb{\Gamma}}$ and $\widetilde{\mathbb{B}}$ are proportional to $\mu$ and therefore vanish in the limit, while $\lim_{\mu\rightarrow 0} \widetilde{\mathbb{A}} = \widetilde{\mathbb{D}}$.
In the zero auxiliary mass limit then, $\hat{\pi}^{\sigma}=0$ and the Hamiltonian of the system becomes
\begin{equation}
H(R, \hat{s}, \pi^{R}, \hat{\pi}^{\sigma} = 0)=\frac{1}{2}\pi^{R\transp}\mathbb{D}\inv\pi^{R} + V(R, \hat{s})
\end{equation}
with $\mathbb{D}\inv=\frac{1}{m_j}\delta_{jj'}$ and where we have used the fact that, see eq.~\ref{eq:MomentaRel}, in the null auxiliary mass limit, $\pi^{\rho}=\mathbb{D}\dot{R}=\pi^R$. Substituting in the expression for the average, we obtain
\begin{equation}
\begin{aligned}\label{eq:ZeroMassMargAverage}
\Braket{\mathcal{O}} =\frac{1}{Z'}\int&\dd^N R\dd^N\pi^{R}O(R, \hat{s}, \pi^{R}, \hat{\pi}^{\sigma}=0)\delta(H(R, \hat{s}, \pi^{R}, \hat{\pi}^{\sigma}=0)-E)\\
\end{aligned}
\end{equation}
The result above implicitly defines the microcanonical marginal probability in the physical phase space in the full adiabatic limit. Interestingly, this definition is in agreement with the form usually assumed for the Born-Oppenheimer probability. The derivation presented here, however, indicates that the dynamical systems rigorously samples this density only in the full $\mu\rightarrow 0$ limit and that, for finite auxiliary masses, corrections, to the mass matrix associated to the momenta would be needed, see eq.~\ref{eq:MomentaRel} and~\ref{eq:AMatrix}.

\section{Conclusions}
In this paper, we have shown how the mass zero constrained dynamics can be adapted to models that impose additional conditions on the dynamics of the auxiliary variables. The interesting case of Born-Oppenheimer \textit{first principles} molecular dynamics for the so-called orbital free approach to DFT was used to illustrate in some detail the extension of the approach for these systems when a single additional condition must be fulfilled. The procedure can be easily extended to the case of multiple additional conditions, thus enabling to tackle also Kohn-Sham DFT propagation. We have also analyzed the statistical mechanics associated to the mass zero constrained dynamics for adiabatically separated systems, deriving the fully adiabatic marginal probability in the physical phase space. This probability coincides with the one typically assumed in Born-Oppenheimer sampling. The discussion presented here focused on the microcanonical ensemble, but the generalization to other ensembles is straightforward. 

\section*{Appendix}
Here we provide the explicit expressions of the block matrices appearing in eq.~\ref{eq:MassMatrix}. Remembering that $\rho=R$ and $\sigma=\sigma(R,s)$, we have
\begin{equation}
\begin{aligned}
\mathbb{A}_{jj'} &= \sum_{i=1}^{3N} m_i \frac{\p R_i}{\p \rho_j}\frac{\p R_i}{\p \rho_{j'}} + \sum_{\alpha=1}^M \mu \frac{\p s_{\alpha}}{\p \rho_j} \frac{\p s_{\alpha}}{\p \rho_{j'}}=  m_j\delta_{ j j'}+ \sum_{\alpha=1}^M \mu \frac{\p s_{\alpha}}{\p \rho_j} \frac{\p s_{\alpha}}{\p \rho_{j'}} \\
&\quad \text{for $ j, j' = 1,\dots, 3N$} \\
\mathbb{B}_{j\beta} &= \sum_{i=1}^{3N} m_i \frac{\p R_i}{\p \rho_j}\frac{\p R_i}{\p \sigma_{\beta}} + \sum_{\alpha=1}^M \mu \frac{\p s_{\alpha}}{\p \rho_j} \frac{\p s_{\alpha}}{\p \sigma_{\beta}}  = \sum_{\alpha=1}^M \mu \frac{\p s_{\alpha}}{\p \rho_j} \frac{\p s_{\alpha}}{\p \sigma_{\beta}} \\
&\quad \text{for $j = 1,\dots, 3N$, $\beta =  1,\dots,M$} \\
\mathbb{\Gamma}_{\beta\beta'} &= \sum_{i=1}^{3N}m_i\frac{\p R_i}{\p \sigma_\beta}\frac{\p R_i}{\p \sigma_{\beta'}} + \sum_{\alpha=1}^{M}\mu \frac{\p s_\alpha}{\p \sigma_\beta}\frac{\p s_\alpha}{\p \sigma_{\beta'}} = \sum_{\alpha=1}^{M}\mu\frac{\p s_\alpha}{\p \sigma_\beta}\frac{\p s_\alpha}{\p \sigma_{\beta'}} \\
&\quad \text{for $\beta, \beta' =  1,\dots,M$} \\
\end{aligned}
\end{equation}
and
\begin{equation}
\begin{aligned}
\mathbb{\Delta}_{jj'} &=\sum_{i=1}^{3N} \frac{1}{m_i} \frac{\p \rho_j}{\p R_i}\frac{\p \rho_{j'}}{\p R_i} \\
&\quad \text{for $ j, j' = 1,\dots, 3N$} \\
\mathbb{E}_{j\beta} &= -\sum_{i=1}^{3N} \frac{1}{m_i} \frac{\p \rho_j}{\p R_i}\frac{\p \sigma_{\beta}}{\p R_i} \\
&\quad \text{for $j = 1,\dots, 3N$, $\beta =  1,\dots,M$} \\
\mathbb{Z}_{\beta\beta'} &= \sum_{i=1}^{3N} \frac{1}{m_i}\frac{\p \sigma_\beta}{\p R_i}\frac{\p \sigma_{\beta'}}{\p R_i} + \sum_{\alpha=1}^{M}\frac{1}{\mu} \frac{\p \sigma_\beta}{\p s_\alpha}\frac{\p \sigma_{\beta'}}{\p s_\alpha} \\
&\quad \text{for $\beta, \beta' =  1,\dots,M$} \\
\end{aligned}
\end{equation}

\section*{Conflicts of interest}
There are no conflicts to declare.

\section*{Acknowledgements}
The authors are grateful to M. Salanne and B. Rothenberg for useful exchanges on the statistics of Born-Oppenheimer dynamics. 





\providecommand*{\mcitethebibliography}{\thebibliography}
\csname @ifundefined\endcsname{endmcitethebibliography}
{\let\endmcitethebibliography\endthebibliography}{}

\end{document}